\documentclass[12pt]{article}
\usepackage{graphicx,amsfonts,amsbsy,times}
\def\id{\mathrm{id}}

\def\ed{\mathrm{d}}

\def\al{\alpha}
\def\be{\beta}
\def\BC{{\mathbb{C}}}

\def\BZ{{\mathbb{Z}}}

\newcommand*{\Ax}{\mathfrak{A}}

\def\AA{{\cal A}}

\def\DD{{\cal D}}

\def\EE{{\cal E}}
\def\LL{{\cal L}}

\def\eq#1\en{\begin{equation}#1\end{equation}}
\def\eqa#1\ena{\begin{eqnarray}#1\end{eqnarray}}
\newcommand*\nn{\nonumber \\}

\newlength{\vscaling} \newlength{\hscaling}
\begin{document}
\begin{titlepage}
\rightline{LMU-TPW 02/05}
\vfill

{\Large\bf
\begin{center}
Noncommutative gerbes and deformation quantization
\end{center}
}

\begin{center}

\vfill

{{\bf Paolo Aschieri, \, Igor Bakovi\' c, \, Branislav Jur\v co, \, Peter Schupp
}}

 \vskip 0.5 cm

Theoretische Physik, Universit\"at M\"unchen\\ 
Theresienstr.\ 37,
80333 M\"unchen, Germany
\end{center}
\vfill
\begin{abstract} 
We define  noncommutative gerbes using the language of
star products. Quantized twisted Poisson structures are
discussed
as an explicit realization in the
sense of deformation quantization.
Our motivation is the noncommutative description of D-branes in the
presence of topologically non-trivial background fields. 
\end{abstract}
\vfill
%\hrule
%\vskip 5pt
%\noindent
%{\footnotesize\it e-mail:
%\parbox[t]{.8\textwidth}{aschieri\,,\,jurco\,,\,schupp\,,%
%\,wess@theorie.physik.uni-muenchen.de}}
\end{titlepage}\vskip.2cm

\newpage

\setcounter{page}{1}
%\thispagestyle{empty}
%%%%%%%%%%%%%%%%%%%%%%%%%%%%%%%%%%%%%%%%%%%%%%%%%%%%%%%%%%%%%%%%%%%%%%%

\section{Introduction}

Gerbes~\cite{Giraud,Brylinski:ab,Hitchin:1999fh} are the next step up from a line bundle on the geometric
ladder in the following sense:
A unitary line bundle is a 1-cocycle in \v Cech cohomology, i.e., it
is a collection of smooth ``transition'' functions
$g_{\alpha\beta}$ on the intersections $U_\alpha \cap U_\beta$ of an 
open cover $\{U_\alpha\}$ of
a manifold $M$ satisfying $g_{\alpha\beta} = - g_{\beta\alpha}$ and
$g_{\alpha\beta}\,g_{\beta\gamma}\,g_{\gamma\alpha}=1$  on
$U_\alpha\cap U_\beta\cap U_\gamma$.
A gerbe is a 2-cocycle in \v Cech cohomology, i.e., it
is a collection $\lambda = \{\lambda_{\alpha\beta\gamma}\}$ of maps 
$\lambda_{\alpha\beta\gamma}:U_\alpha\cap U_\beta\cap U_\gamma
\to U(1)$,  valued in the abelian group $U(1)$, satisfying
\eq
\lambda_{\alpha\beta\gamma}=
\lambda^{-1}_{\beta\alpha\gamma}=
\lambda^{-1}_{\alpha\gamma\beta}=
\lambda^{-1}_{\gamma\beta\alpha}
\en 
and the 2-cocycle condition
\eq
\delta \lambda=\lambda_{\beta\gamma\delta} \,
\lambda^{-1}_{\alpha\gamma\delta}\,\lambda_{\alpha\beta\delta}\,
  \lambda^{-1}_{\alpha\beta\gamma}=1
\en
on $U_\alpha\cap U_\beta\cap U_\gamma \cap U_\delta$.
The collection $\lambda = \{\lambda_{\alpha\beta\gamma}\}$ of maps 
with the stated properties defines a gerbe in the
same sense as a collection of transition functions defines
a line bundle.
In the special case where $\lambda$ is a \v Cech 2-coboundary with 
$\lambda = \delta h$, i.e.,
$\lambda_{\alpha\beta\gamma} =  
h_{\alpha\beta}\,h_{\beta\gamma}\,h_{\gamma\alpha}$, we call the
collection $h = \{ h_{\alpha\beta} \}$ of functions 
$h_{\alpha\beta} : U_\alpha \cap U_\beta \to U(1)$
a trivialization of a gerbe. Taking the ``difference'' 
of two trivializations
$\{h_{\alpha\beta}\}$, $\{h'_{\alpha\beta}\}$ of a gerbe 
we step down the geometric ladder again
and obtain a line bundle: 
$g_{\alpha\beta} \equiv h_{\alpha\beta} /h'_{\alpha\beta}$
satisfies the 1-cocycle condition 
$g_{\alpha\beta}\,g_{\beta\gamma}\,g_{\gamma\alpha}=1$.

A gerbe has a local trivialization for any particular 
open set $U_0$ of the covering:
Defining $h_{\beta\gamma}\equiv\lambda_{0\beta\gamma}$ with 
$\beta,\gamma \neq 0$  we find from the 2-cocycle condition of 
a gerbe that $\lambda_{\alpha\beta\gamma} =  
h_{\alpha\beta}\,h_{\beta\gamma}\,h_{\gamma\alpha}$. 
This observation
leads to an equivalent definition of a gerbe in terms of line bundles
on the double overlaps of the cover.
The only difference to the definition of a 
line bundle from this point of view is that we step up the
geometric ladder and use line bundles on 
$U_{\alpha}\cap U_{\beta}$ rather than
transition functions. A gerbe is then a collection of
line bundles $L_{\alpha\beta}$ for each 
double overlap $U_\alpha\cap U_\beta$, such that:
\begin{description}
\item[G1] There is an isomorphism 
$L_{\alpha\beta} \cong L^{-1}_{\beta\alpha}$.
\item[G2] There is a trivialization 
$\lambda_{\alpha\beta\gamma}$  of 
$L_{\alpha\beta}\otimes L_{\beta\gamma} \otimes L_{\gamma\alpha}$
 on $U_{\alpha}\cap U_{\beta}\cap U_{\gamma}$.
\item[G3] The trivialization 
$\lambda_{\alpha\beta\gamma}$ satisfies $\delta \lambda=1$
on $U_{\alpha}\cap U_{\beta}\cap U_{\gamma}\cap U_{\delta}$.
\end{description}

Gerbes are interesting in physics for several reasons:
One motivation is the interpretation of D-brane charges
in terms of K-theory in the presence of a topologically
nontrivial $B$-field, when the gauge fields living on D-branes
become connections on certain noncommutative algebras rather
than on a vector 
bundle~\cite{Witten:1998cd}-\cite{Carey:2002xp}. 
Azumaya algebras appear to be a 
natural choice and give the link to gerbes.
Gerbes, rather than line bundles, are the structure
that arises in the presence of closed 3-form backgrounds
as, e.g., in WZW models 
and Poisson sigma models with WZW 
term~\cite{Park:2000au,Klimcik:2001vg,Gawedzki:2002se}.  
Gerbes help illuminate the geometry of mirror symmetry of 
3-dimensional Calabi-Yau manifolds~\cite{Hitchin:1999fh} and they
provide a language to formulate duality transformations
with higher order antisymmetric fields~\cite{Caicedo:2002qd}.
Our motivation is the noncommutative description of D-branes in the
presence of topologically non-trivial background fields.

The paper is organised as follows: 
In section~2 we recall the local description of noncommutative
line bundles in the framework of deformation quantization. 
Instead of repeating that construction we shall take 
the properties that were derived in~\cite{Jurco:2001kp,Jurco:2000fs}
as a formal definition of a noncommutative line bundle.
In the same spirit we define noncommutative gerbes in 
section~3 using the language
of star products and complement this definition
with an explicit realization of  noncommutative gerbes
as quantizations of twisted Poisson structures as
introduced in \cite{Severa:2001qm} and further discussed 
in~\cite{Severa2}.

\section{Noncommutative line bundles}
\label{sec2}

Here we collect some facts on noncommutative line 
bundles~\cite{b,Jurco:2001kp} that we will need 
in the sequel.\footnote{A 
noncommutative line bundle is a finite projective 
module. In  the present context it can be understood as a quantization
of a line bundle in the sense of deformation quantization.
Here we shall take the properties of quantized line bundles
as derived in~\cite{Jurco:2001kp,Jurco:2000fs} as a formal definition
of a noncommutative line bundle.} 
Let $(M, \theta)$ be a general Poisson manifold, and $\star$ the
corresponding
Kontsevich's deformation quantization of the Poisson tensor $\theta$.
Further let us consider 
a good covering  $\{ U^i\}$ of $M$. For purposes of this paper a 
noncommutative line bundle 
$\cal L$ is defined by
a collection of local transition functions $G^{ij} \in 
C^\infty(U^i \cap
U^j)[[\hbar]]$, valued in the enveloping
algebra of $U(1)$ (see \cite{Jurco:2000ja}),
and a collection of maps $\DD^i : C^\infty
(U^i)[[\hbar]]\rightarrow
C^\infty(U^i)[[\hbar]]$, formal power series in $\hbar$ starting with
identity and with coefficients
being differential operators such that 
\eq
G^{ij} \star G^{jk} = G^{ik} \label{transition}
\en
on $U^i\cap U^j \cap U^k$, $G^{ii} ={} 1$ on $U^i$, and
\eq
\mathrm{Ad}_\star G^{ij} ={} \DD^i \circ (\DD^{j})
^{-1} \label{covariance}
\en
on $U^i\cap U^j$ or, equivalently, 
$\DD^i(f) \star G^{ij} ={} G^{ij} \star \DD^j(f)$
for all $f \in C^\infty(U^i \cap
U^j)[[\hbar]]$.
Obviously, with this definition the local maps $\DD^i$ can be used to
define \emph{globally} 
a new star product $\star'$ (because the inner automorphisms
$\mathrm{Ad}_\star G^{ij}$ do not affect $\star'$)
\eq
\DD^i(f\star' g) = \DD^i f \star \DD^i g \;. \label{starprime}
\en
We say that two line bundles ${\cal L}_1={}\{G_1^{ij}, \DD_1^i, \star\}$ and 
${\cal L}_2={}\{G_2^{ij},\DD_2^i, \star\}$ are equivalent if 
there exist
a collection of invertible local functions $H^i\in C^\infty
(U^i)[[\hbar]]$ such that
\eq 
G_1^{ij} = H^i\star G_2^{ij}\star (H^j)^{-1}
\en
and
\eq
\DD_1^i = \mathrm{Ad}_\star H^i \circ  \DD_2^i\;.
%\DD_1^i(.) ={} H^i \star \DD_2^i(.) \star (H^i)^{-1}.
\en
The tensor product of two line bundles 
${\cal L}_1=\{G^{ij}_1, \DD_1^i,
\star_1\}$ and
${\cal L}_2=\{G_2^{ij}, \DD_2^i, \star_2\}$
is well defined if $\star_2 = \star'_1$ (or $\star_1 = \star'_2$.) Then the corresponding tensor
product is a line bundle
${\cal L}_2\otimes {\cal L}_1 ={\cal L}_{21}={}\{G_{12}^{ij},
\DD_{12}^{ij},\star_1\}$ defined as
\eq
G_{12}^{ij} = \DD_1^i(G_2^{ij})\star_1 G_1^{ij}= G_1^{ij}\star_1 
\DD_1^j(G_2^{ij}) \label{tensora}
\en
and
\eq
\DD_{12}^{i}= \DD_1^{i}\circ\DD_2^{i} \;. \label{tensorb}
\en
The order of indices of 
${\cal L}_{21}$ indicates the bimodule structure of the corresponding
space of sections to be defined later, whereas the first index on the
$G_{12}$'s and $\DD_{12}$'s
indicates the star product (here: $\star_1$) 
by which the objects multiply.

A section $\Psi ={}(\Psi^i)$ is a collection of functions 
$\Psi^i \in C_{\BC}^\infty(U^i)[[\hbar]]$ satisfying consistency
relations
\eq
\Psi^i = G^{ij} \star \Psi^i \label{section}
\en
on all intersections
$U^i \cap U^j$. With this definition the space of sections
$\EE$ is a right $\Ax{}=(C^\infty(M)[[\hbar]],\star)$ module. We
shall use the notation
${\cal E}_{\Ax}$ for it. The right action of the function $f\in \Ax$
is the regular one
\eq
\Psi. f = (\Psi^k \star f)\;.
\en
Using the maps $\DD^i$ it is easy to turn ${\cal E}$ also into a left
${\Ax}' ={}(C^\infty(M)[[\hbar]],\star')$ 
module ${}_{{\Ax}'}{\cal E}$. The left action of ${\Ax}'$ is given by
\eq  
f .\Psi = (\DD^i(f) \star \Psi^i)\;. \label{leftaction}
\en
It is easy to check, using  (\ref{covariance}), that the left action
(\ref{leftaction}) is
compatible with (\ref{section}). From  the property (\ref{starprime})
of the maps $\DD^i$ we find 
\eq
f.(g.\Psi) = (f \star' g).\Psi \;.
\en 
Together we have a bimodule structure ${}_{{\Ax}'}{\cal E}_{\Ax}$
on the space of sections. 
There is an obvious way of tensoring sections. The section
\eq 
\Psi_{12}^i = \DD_1^i (\Psi_2^i) \star_1 \Psi_1^i \label{tensorc}
\en
is a section of the tensor product line bundle (\ref{tensora}), 
(\ref{tensorb}).
Tensoring of line bundles naturally corresponds to tensoring of 
bimodules.

Using the Hochschild complex we can introduce a natural differential
calculus on the algebra $\Ax$.\footnote{Other choices for the
differential calculus are of course possible,
e.g., the Lie algebra complex.}  The $p$-cochains, elements of 
$C^p={}\mbox{Hom}_{\BC}({\Ax}^{\otimes p},\Ax)$,
play the role of $p$-forms and the derivation 
$\ed: C^p \rightarrow C^{p+1}$ is given on 
$C\in C^p$ as
\eqa
\lefteqn{\ed C_{\,}(f_1,f_2,\ldots,f_{p+1})  ={}  f_1 \star 
C(f_2,\ldots,f_{p+1}) - 
C(f_1\star f_2,\dots,f_{p+1})}\nn
&& {} + C(f_1,f_2\star
f_3,\ldots,f_{p+1})-\ldots+(-1)^p C(f_1,f_2,\ldots,f_p\star f_{p+1})\nn
&& {} +(-1)^{p+1} C(f_1,f_2,\ldots,f_p)\star f_{p+1}\;. \label{der}
\ena
A (contravariant) connection $\nabla : {\cal E}\otimes_{\Ax} C^p 
\rightarrow {\cal E}\otimes_{\Ax} C^{p+1}$ can now be defined by a 
formula
similar to (\ref{der}) using the natural extension of the
left and right module structure of $\EE$ to ${\cal E}\otimes_{\Ax}  
C^p$. 
Namely, for a $\Phi\in {\cal E}\otimes_{\Ax} C^p$ we have
\eqa
\lefteqn{\nabla \Phi_{\,}(f_1,f_2,\ldots,f_{p+1})  =
f_1.\Phi(f_2,\ldots,f_{p+1}) - 
\Phi(f_1\star f_2,\ldots,f_{p+1})} \nn
&& {} + \Phi(f_1,f_2\star
f_3,\ldots,f_{p+1})-\ldots+(-1)^p\Phi(f_1,f_2,\ldots,f_p\star f_{p+1}) 
\nn
&& {} +(-1)^{p+1} \Phi(f_1,f_2,\ldots,f_p).f_{p+1}\;. \label{c}
\ena
We also have the cup product $C_1 \cup C_2 $ of two cochains 
$C_1\in C^{p}$ and $C_2\in C^{q}$;
\eq
(C_1\cup C_2)(f_1,...,f_{p+q})={}C_1(f_1,...,f_p)\star
C_2(f_{p+1},...,f_q)\;.
\en
The cup product extends to a map from $({\cal E}
\otimes_{\Ax} C^p) \otimes_{\Ax} C^q$ to ${\cal E} \otimes_{\Ax} 
C^{p+q}.$ The connection
$\nabla$ satisfies the graded Leibniz rule
with respect to the cup product and thus 
defines a bona fide connection on the module ${\cal E}_{\Ax}$.
On the sections the connection $\nabla$ introduced here is simply the
difference between the
two actions of $C^{\infty}(M)[[\hbar]]$ on $\EE$:
\eq
\nabla \Psi_{\,}(f) ={} f.\Psi - \Psi.f =\left(\nabla^i \Psi^i (f)\right)
= \left(\DD^i(f)\star\Psi^i - \Psi^i \star f\right)\;.\label{cprime}
\en
As in \cite{Jurco:2000fs} 
we define the gauge potential ${\cal A}=({\cal A}^i)$, where the
${\cal A}^i\!: C^\infty(U^i)[[\hbar]] 
\rightarrow C^\infty(U^i)[[\hbar]]$ are 
local 1-cochains,  by
\eq
{\cal A}{^i}\equiv {\cal D}{^i} - \id \label{Dminusid}~.
\en
Then we have for a section $\Psi = (\Psi^i)$,
where the 
$\Psi^i \in C_{\BC}^\infty(U^i)[[\hbar]]$ are local $0$-cochains,
\eq
\nabla^i\Psi^i{\,}(f)= \ed\Psi^i_{\,\;}(f)_{\,}+_{\,}\AA^i(f)\star\Psi^i \;,
\label{ndA}
\en
and more generally $\,\nabla^i\Phi^i= \ed\Phi^i+\AA^i\cup\Phi^i$ with 
$\Phi=(\Phi^i)\in {\cal E}\otimes_{\Ax} C^p$.
In the intersections $U^i \cap U^j$ we have 
the gauge transformation (cf. (\ref{covariance}))
\eq
{\cal A}^i={\rm Ad}_\star G^{ij} \circ {\cal A}^j 
+G^{ij}\star \ed(G^{ij})^{-1}~. 
\label{gaugetransfA}
\en
The curvature $K_\nabla \equiv \nabla^2 :{\cal E}\otimes_{\Ax} C^p 
\rightarrow {\cal
E}\otimes_{\Ax} C^{p+2}$  corresponding to the connection $\nabla$, 
measures the difference between the two star products $\star'$ and
$\star$. On a section $\Psi$, it is given by
\eq
(K_\nabla \Psi)(f,g)= \left(\DD^i(f\star'g- f\star g)\star\Psi^i\right)
\;.\label{curv} 
\en
The connection for the tensor product line bundle (\ref{tensora})
is given on sections as
\eq
\nabla_{12} \Psi_{12}^i = \DD^i_1 (\nabla_2 \Psi_2^i)\star_1 \Psi_1^i +
\DD^i_1(\Psi_2)\star_1\nabla_1\Psi^i_1\;. 
\en
Symbolically,
\eq
\nabla_{12} = \nabla_1 + \DD_1 (\nabla_2) \label{tensorconnection}.
\en
Let us note that 
the space of sections $\EE$  as a right $\Ax$-module
is projective of finite type. Of course, the same holds if  $\EE$ is 
considered
as a left $\Ax'$ module. Also let us note that the two algebras $\Ax$ and
$\Ax'$ are Morita equivalent. Up to a global isomorphism they must be 
related
by an action of the Picard group $\mbox{Pic}(M) \cong H^2(M,\mathbb{Z})$
as follows.
Let $L\in \mbox{Pic}(M)$ be a (complex) line bundle on $M$ and $F$ its 
Chern class. 
Consider the formal Poisson structure
$\theta'$ given by the geometric series
\eq 
\theta'= \theta (1+ \hbar F\theta)^{-1}.
\en
In this formula $\theta$ and $F$ are understood as maps $\theta: T^*M 
\rightarrow  TM$, 
$F:TM \rightarrow  T^*M$ and  
$\theta'$ is the result of the indicated map compositions. Then $\star'$ 
must  (up to a global isomorphism) be
the deformation quantization of $\theta'$ corresponding to some 
$F\in H^2(M,\mathbb{Z})$. 
If $F={}da$  then the corresponding quantum line bundle is trivial, i.e.,
\eq
G^{ij} ={} (H^i)^{-1} \star H^j
\en
and the linear map  
\eq
\DD = \mathrm{Ad}_\star H^i \circ \DD^i 
%\DD(.)=H^i \star \DD^i(.) \star (H^i)^{-1}
\en
defines a global equivalence  (a stronger notion than 
Morita equivalence) of $\star$ and $\star'$.

\section{Noncommutative gerbes}

Now let us consider any covering $\{U_\alpha\}$ (not necessarily a good 
one) of a 
manifold
$M$. Here we switch from upper Latin to lower Greek indices to label 
the 
local patches. The reason for
the different notation will become clear soon. Consider each local patch
equipped with its own star product $\star_\alpha$ the deformation 
quantization
of a local Poisson structure $\theta_\alpha$. We assume that on each 
double
intersection $U_{\alpha \beta}={}U_\alpha \cap U_\beta$ the local 
Poisson structures $\theta_\alpha$
and $\theta_\beta$ are related
similarly as in the previous section via some integral closed two form 
$F_{\beta \alpha}$, which is
the curvature of a line bundle $L_{\beta \alpha}\in \mbox{Pic}(U_{\alpha 
\beta})$
\eq
\theta_\alpha ={} \theta_\beta(1+\hbar F_{\beta \alpha}\theta_\beta)^{-1}.\label{pis}
\en

\begin{figure}[tb]
\centerline{\includegraphics{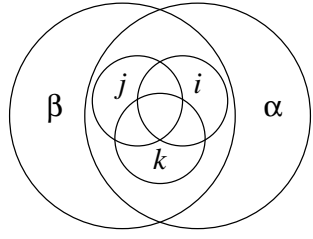}}
\caption{\small Double intersection $U_\alpha \cap U_\beta$
equipped with a NC line bundle $G_{\alpha \beta}^{ij} 
\star_\alpha G_{\alpha \beta}^{jk} = G_{\alpha
\beta}^{ik}$.} \label{ger2}
\end{figure}
Let us now consider a good covering
$U^i_{\alpha\beta}$ of each double intersection 
$U_\alpha \cap U_\beta$  with a noncommutative line bundle
${\cal L}_{\beta \alpha}={}\{G_{\alpha \beta}^{ij}, \DD_{\alpha \beta}^i,
\star_\alpha\}$, see Figure~\ref{ger2},
\eq
G_{\alpha \beta}^{ij} \star_\alpha G_{\alpha \beta}^{jk} = G_{\alpha
\beta}^{ik}\;, \qquad G_{\alpha \beta}^{ii} = 1\;, \label{transitiong}
\en
\eq
\DD_{\alpha \beta}^i(f) \star_\alpha G_{\alpha \beta}^{ij}  =
G^{ij}_{\alpha \beta} \star_\alpha \DD_{\alpha
\beta}^j(f)\label{covarianceg}
\en
and
\eq
\DD_{\alpha \beta}^i(f\star_\beta g) = \DD_{\alpha \beta}^i (f) 
\star_\alpha
\DD_{\alpha \beta}^i (g)\;. \label{starprimeg}
\en
The opposite order of indices labelling the line bundles and the
corresponding transition functions and equivalences
simply reflects a choice of convention. 
%This is the result 
%of maybe not the best choice of conventions.  
As in the previous section 
the order of indices of 
${\cal L}_{\alpha \beta}$ indicates the bimodule structure of the corresponding
space of sections, whereas the order of Greek indices on $G$'s and $D$'s
indicates the star product in which the objects multiply.
The product always goes with the first index of the multiplied objects.

A noncommutative gerbe is characterised by the following axioms:
\begin{description} 
\item[Axiom 1]
${\cal L}_{\alpha \beta}= \{G_{\beta \alpha}^{ij}, \DD_{\beta \alpha}^i,
\star_\beta\}$ and ${\cal L}_{\beta \alpha}=\{G_{\alpha \beta}^{ij}, \DD_{\alpha \beta}^i,
\star_\alpha\}$ are related as 
follows
\eq
\{G_{\beta \alpha}^{ij}, \DD_{\beta \alpha}^i,
\star_\beta\}={}\{(\DD_{\alpha \beta}^j)^{-1}(G_{\alpha \beta}^{ji}),
(\DD^i_{\alpha \beta})^{-1},
\star_\beta\}\label{inverse}
\en
i.e. $\LL_{\al\be}=\LL_{\be\al}^{-1}$. (Notice also that 
$(\DD_{\alpha \beta}^j)^{-1}(G_{\alpha \beta}^{ji})=
(\DD_{\alpha \beta}^i)^{-1}(G_{\alpha \beta}^{ji})\,$.)

\item[Axiom 2]
On the triple intersection 
$U_{\alpha}\cap U_{\beta}\cap U_{\gamma}$
the tensor product 
${\cal L}_{\gamma \beta} \otimes {\cal L}_{\beta \alpha}$ 
is equivalent to the line bundle ${\cal L}_{\gamma \alpha}$ . 
Explicitly
\eq
G_{\alpha \beta}^{ij}\star_\alpha\DD^j_{\alpha \beta}(G_{\beta 
\gamma}^{ij})=
\Lambda^i_{\alpha \beta \gamma} \star_{\alpha}G_{\alpha 
\gamma}^{ij}\star_\alpha
(\Lambda^j)^{-1}_{\alpha \beta \gamma}\;,\label{tripletensorG}
\en
\eq
\DD^i_{\alpha \beta}\circ \DD^i_{ \beta \gamma}
= \mathrm{Ad}_{\star_\alpha} \Lambda^i_{\alpha \beta \gamma}
\circ \DD^i_{\alpha \gamma} \;.\label{tripletensorD}
%\DD^i_{\alpha \beta}\circ \DD^i_{ \beta \gamma}(f) 
%= \Lambda^i_{\alpha \beta
%\gamma}\star_\alpha
%\DD^i_{\alpha \gamma}(f) \star_\alpha
%(\Lambda^i)^{-1}_{\alpha \beta \gamma}\;.
\en 

\item[Axiom 3]
On the quadruple intersection 
$U_{\alpha}\cap U_{\beta} \cap U_{\gamma} \cap U_{\delta}$
\eq
\Lambda^i_{\alpha \beta \gamma}\star_\alpha \Lambda^i_{\alpha \gamma 
\delta}=
\DD^i_{\alpha \beta}(\Lambda^i_{\beta \gamma \delta})\star_\alpha 
\Lambda^i_{\alpha \beta \delta}\label{gerbe}\;,
\en
\eq 
\Lambda^i_{\alpha \beta \gamma}=(\Lambda^i_{\alpha \gamma \beta})^{-1} 
\hskip 1cm \mbox{and}\hskip 1cm
\DD^i_{\alpha \beta}(\Lambda^i_{\beta \gamma \alpha}) =
\Lambda^i_{\alpha \beta \gamma}\label{inversegerbe}\;.
\en
\end{description}
With slight abuse of notation we have used  Latin indices $\{i,j,..\}$ 
to label both the good coverings
of the intersection of the local patches $U_\alpha$ 
and the corresponding transition 
functions of the consistent restrictions of line bundles 
${\cal L}_{\alpha \beta}$ 
to these intersections.
A short comment on the consistency of Axiom~3 is in order.
Let us define 
\eq
\DD^i_{\alpha \beta \gamma}=\DD^i_{\alpha \beta} \circ \DD^i_{\beta
\gamma}\circ \DD^i_{\gamma \alpha}\;.
\en
Then it is easy to see that
\eq
\DD^i_{\alpha \beta \gamma}
\circ \DD^i_{\alpha \gamma \delta}\circ \DD^i_{\alpha 
\delta \beta }=
\DD^i_{\alpha \beta}\circ \DD^i_{\beta \gamma \delta}
\circ \DD^i_{\beta \alpha}\;.
\en 
In view of (\ref{tripletensorD}) this implies that
$$ \Lambda^i_{\alpha \beta \gamma \delta}\equiv\DD^i_{\alpha
\beta}(\Lambda^i_{\beta \gamma \delta})\star_\alpha 
\Lambda^i_{\alpha \beta \delta}\star_\alpha \Lambda^i_{\alpha \delta
\gamma}\star_\alpha 
\Lambda^i_{\alpha \gamma \beta}$$ is central. Using this and the
associativity of 
$\star_{\alpha}$ together
with (\ref{tripletensorG})
applied to the triple tensor product ${\cal L}_{\delta \gamma}
\otimes {\cal L}_{\gamma \beta} \otimes {\cal L}_{\beta \alpha}$
transition functions
\eq
G^{ij}_{\alpha \beta \gamma}\equiv G_{\alpha 
\beta}^{ij}\star_\alpha\DD^j_{\alpha \beta}
(G_{\beta \gamma}^{ij})\star_\alpha \DD^j_{\alpha \beta}(\DD^j_{\beta
\gamma}
(G_{\gamma \delta}^{ij}))\label{tripletr}
\en 
reveals that $\Lambda^i_{\alpha \beta \gamma \delta}$ is independent of 
$i$.
It is therefore consistent to set $\Lambda^i_{\alpha \beta \gamma \delta}$ 
equal to~$1$.
A similar consistency check works also for (\ref{inversegerbe}).
If we replace all noncommutative line bundles
$\mathcal{L}_{\alpha\beta}$ in Axioms~1-3 by equivalent ones, we get
by definition an equivalent noncommutative gerbe.

There is a natural (contravariant) connection on a quantum gerbe. It is 
defined using
the (contravariant) connections 
$\nabla_{\alpha \beta}=(\nabla_{\alpha \beta}^i)$
(cf. (\ref{c}), (\ref{cprime})) on quantum line bundles 
${\cal L}_{\beta \alpha}$. Let us denote by 
$\nabla_{\alpha \beta \gamma}$ the contravariant
connection formed on the triple tensor product 
${\cal L}_{\alpha \gamma \beta} \equiv{\cal L}_{\alpha \gamma} 
\otimes{\cal L}_{\gamma \beta} 
\otimes {\cal L}_{\beta \alpha}$ with maps 
$\DD^i_{\alpha \beta \gamma}$ and transition functions 
(\ref{tripletr}) 
according to the rule (\ref{tensorconnection}). 
Axiom~2 states that $\Lambda_{\alpha \beta \gamma}^i$ is a 
trivialization of ${\cal L}_{\alpha \gamma \beta}$ and that
\eq
\nabla_{\alpha \beta \gamma}^i \Lambda^{i}_{\alpha \beta \gamma}=0~.
\en
Using Axiom~2 one can show that the product bundle
\eq
{\cal L}_{\alpha \beta \gamma \delta}={\cal L}_{\alpha \beta \gamma}\otimes {\cal L}_{\alpha \gamma \delta}
\otimes{\cal L}_{\alpha \delta \beta}\otimes{\cal L}_{\alpha \beta}\otimes
{\cal L}_{\beta \delta \gamma}\otimes{\cal L}_{\beta \alpha}
\en
is trivial: it has transition functions 
$G^{ij}_{\alpha \beta \gamma \delta}=1$ and maps
$\DD^i_{\alpha \beta \gamma \delta}=\id$. 
The constant unit section is thus well defined on this bundle.
On ${\cal L}_{\alpha \beta \gamma \delta}$ 
we also have the section $(\Lambda^i_{\alpha \beta \gamma \delta})$.
Axiom~3 implies 
$({\Lambda}^i_{\alpha \beta \gamma \delta})$ to be the unit section. 
If two of the indices $ \alpha,\,
\beta,\, \gamma,\, \delta $ are equal, triviality of 
the bundle ${\cal L}_{\alpha \beta \gamma \delta}$
implies (\ref{inversegerbe}).
Using for example the first relation in (\ref{inversegerbe})
one can show that (\ref{gerbe}) written in the form
$
\,\DD^i_{\alpha \beta}(\Lambda^i_{\beta \gamma \delta})\star_\alpha 
\Lambda^i_{\alpha \beta \delta\,} \star_\alpha
 \Lambda^i_{\alpha \delta  \gamma\,}\star_\alpha
\Lambda^i_{\alpha \gamma\beta }=1\,
$
is invariant under cyclic permutations of any three of the 
four factors appearing on the l.h.s..

If we now assume that 
$F_{\alpha \beta}={}da_{\alpha \beta}$ for each $U_{\alpha}\cap U_{\beta}$
then all line bundles
${\cal L}_{ \beta \alpha}$ are trivial
$$
G^{ij}_{\alpha \beta} = (H_{\alpha \beta}^i)^{-1}\star_\alpha
H_{\alpha \beta}^j
$$
$$
\DD_{\alpha \beta} = \mathrm{Ad}_{\star_\alpha} H_{\alpha \beta}^i
\circ \DD^i_{\alpha \beta} \; .
%\DD^i_{\alpha \beta}(.) ={} (H_{\alpha \beta}^i)^{-1}\star_\alpha
%\DD_{\alpha \beta}(.)\star_\alpha H_{\alpha \beta}^i \;,
$$
It then easily follows that
\eq 
\Lambda_{\alpha \beta \gamma} \equiv H^i_{\alpha \beta}\star_\alpha 
\DD^i_{\alpha \beta}
(H^i_{\beta \gamma}) \star_\alpha 
\DD^i_{\alpha \beta}\DD^i_{\beta\gamma}(H^i_{\gamma
\alpha})\star_\alpha\Lambda^i_{\alpha \beta \gamma} 
\en 
defines a global function on the triple intersection $U_{\alpha}\cap 
U_{\beta}\cap
U_{\gamma}$. 
$\Lambda_{\alpha \beta \gamma}$ is just the quotient of the two sections
$\big(H^i_{\alpha \beta}\star_\alpha 
\DD^i_{\alpha \beta}
(H^i_{\beta \gamma})\star_\alpha 
\DD^i_{\alpha \beta}\DD^i_{\beta\gamma}(H^i_{\gamma
\alpha})\big)^{-1}$ and 
$\Lambda^i_{\alpha \beta \gamma}$ of the triple tensor product
${\cal L}_{\alpha \gamma} 
\otimes{\cal L}_{\gamma \beta} 
\otimes {\cal L}_{\beta \alpha}$.
On the quadruple overlap $U_{\alpha}\cap U_{\beta} \cap U_{\gamma} \cap 
U_{\delta}$ it satisfies
conditions analogous to (\ref{gerbe}) and (\ref{inversegerbe})
\eq
\Lambda_{\alpha \beta \gamma}\star_\alpha \Lambda_{\alpha \gamma \delta}=
\DD_{\alpha \beta}(\Lambda_{\beta \gamma \delta})\star_\alpha 
\Lambda_{\alpha \beta \delta}
\label{Gerbe}\;,
\en
\eq 
\Lambda_{\alpha \beta \gamma}=(\Lambda_{\alpha \gamma 
\beta})^{-1}\hskip 1cm \mbox{and} \hskip 1cm
\DD_{\alpha \beta}(\Lambda_{\beta \gamma \alpha})= \Lambda_{\alpha 
\beta \gamma}
\label{Inversegerbe}\;.
\en
Also
\eq
\DD_{\alpha \beta}\circ\DD_{\beta \gamma}\circ\DD_{\gamma \alpha}=
\mathrm{Ad}_{\star_\alpha}\Lambda_{\alpha \beta \gamma}\;
\label{stack}.
\en
So we can take formulas (\ref{Gerbe})-%, (\ref{Inversegerbe}),
(\ref{stack}) 
as a definition of a gerbe
in the case of a good covering $\{U_\alpha\}$.
The collection of local equivalences $\DD_{\alpha \beta}$ satisfying
(\ref{stack}) with $\Lambda_{\alpha \beta \gamma}$ fulfilling 
(\ref{Gerbe}),
(\ref{Inversegerbe}) defines on $M$ a stack of algebras \cite{Kashiwara}.

{}From now on we shall consider only good coverings.
A noncommutative gerbe defined by $\Lambda_{\alpha \beta \gamma}$  
and $\DD_{\alpha\beta}$
is said to be trivial if there exist a global star product
$\star$ on $M$ and a collection of ``twisted'' transition functions
$G_{\alpha \beta}$ defined on each overlap $U_{\alpha}\cap U_{\beta}$
and a collection $\DD_\alpha$ of local equivalences between the 
global  product
$\star$ and the local products~$\star_\alpha$
$$\DD_{\alpha}(f)\star \DD_{\alpha}(g) ={} \DD_{\alpha}(f\star_\alpha
g)$$
satisfying the following two conditions:
\eq G_{\alpha \beta }\star G_{\beta \gamma }  =
\DD_\alpha(\Lambda_{\alpha \beta \gamma})\star 
G_{\alpha \gamma}
\en
and
\eq 
\mathrm{Ad}_\star G_{\alpha\beta} \circ \DD_{\beta}
= \DD_{\alpha} \circ \DD_{\alpha\beta}\;.\label{GDDG}
%G_{\alpha \beta }\star \DD_{\beta}(.) ={} \DD_{\alpha}\DD_{\alpha
%\beta}(.) \star G_{\alpha \beta}. 
\en
Locally, every noncommutative gerbe is trivial as is easily seen from 
(\ref{Gerbe}), (\ref{Inversegerbe}) and
(\ref{stack}) by fixing the index $\alpha$.
Defining as in (\ref{Dminusid}), ${\cal A}_\al={\cal D}{_\al}-\id$,
 ${\cal A}_{\al\be}={\cal D}_{\al\be}-\id$ we obtain  
the ``twisted'' gauge transformations  
\eq
{\cal A}_\al={\rm Ad}_\star G_{\al\be} \circ {\cal A}_{\be} 
+G_{\al\be}\star \ed(G_{\al\be})^{-1} 
-{\cal D}_\al \circ {\cal A}_{\al\be} \;.
\en

\section{Quantization of twisted Poisson structures}

Let $H\in H^3(M,\BZ)$ be a closed integral three form on M. 
Such a form is known 
to define a gerbe on M. We can find a good covering $\{U_{\alpha}\}$
and local potentials $B_{\alpha}$ with $H=dB_{\alpha}$ for $H$. 
On $U_{\alpha} \cap U_{\beta}$ the difference of the two local potentials
$B_{\alpha} - B_{\beta}$ is closed and hence exact: $B_{\alpha} - B_{\beta}= da_{\alpha \beta}$.
On a triple intersection $U_\alpha \cap U_\beta \cap U_\gamma$
we have
\eq
a_{\alpha \beta} + a_{\beta \gamma} + a_{\gamma \alpha} 
= -i\lambda_{\alpha \beta \gamma}
d\lambda_{\alpha \beta \gamma}^{-1}.
\en
The collection of local functions 
$\lambda_{\alpha \beta \gamma}$ defines the gerbe. 

Let us also assume the existence
of a formal antisymmetric bivector field $\theta= \theta^{(0)} + \hbar 	
\theta^{(1)}+\ldots$
on $M$ such that
\eq
[\theta, \theta]= \hbar\; \theta^{*\!}H \;, \label{twistpoisson}
\en
where $[\;,\,]$ is the Schouten-Nijenhuis bracket and $\theta^*$ denotes the natural map 
sending $n$-forms to $n$-vector fields by 
``using $\theta$ to raise indices''. Explicitly,
in local coordinates, 
$\theta^*H^{ijk} = \theta^{im}\theta^{jn}\theta^{ko} H_{mno}$.
We call $\theta$  a Poisson structure twisted by $H$
\cite{Severa:2001qm,Park:2000au,Klimcik:2001vg}. 
On each $U_{\alpha}$ we can introduce a local formal Poisson structure
$\theta_\alpha = \theta(1-\hbar B_{\alpha}\theta)^{-1}$, 
$[\theta_\alpha, \theta_\alpha]=0$.
The Poisson structures $\theta_\alpha$ 
and $\theta_\beta$
are related on the intersection $U_\alpha \cap U_\beta$ as in (\ref{pis})
\eq
\theta_\alpha = \theta_\beta(1+\hbar F_{\beta \alpha}\theta_\beta)^{-1} \;,
\en
with an exact $F_{\beta \alpha}= d a_{\beta \alpha}$.
Now we can use Formality \cite{Kontsevich:1997vb} 
to obtain local star products $\star_{\alpha}$ and 
to construct for each intersection $U_\alpha \cap U_\beta$ 
the corresponding equivalence
maps $\DD_{\alpha \beta}$. See \cite{Jurco:2000fs,Jurco:2001kp} 
for an explicit formula for
the equivalence maps. According to our discussion 
in the previous section
these $\DD_{\alpha \beta}$, supplemented by trivial
transition functions, define a collection of trivial line bundles 
${\cal L}_{\beta \alpha}$.
On each triple intersection we then have
\eq
\DD_{\alpha \beta}\circ\DD_{\beta \gamma}\circ\DD_{\gamma \alpha}
= \mathrm{Ad}_{\star_\alpha}\Lambda_{\alpha \beta \gamma}\;.
\en
It follows from the discussion after formula (\ref{inversegerbe}) that
$\Lambda_{\alpha \beta \gamma}$ defines a quantum
gerbe (a deformation quantization of the classical gerbe $\lambda_{\alpha \beta \gamma}$)
if each of the central functions $\Lambda_{\alpha \beta \gamma \delta}$
introduced there can be chosen to be equal to $1$. 
See  \cite[section~5]{Severa2} and \cite{Kontsevich2} 
that this is really the case.

\subsection*{Acknowledgement}

B.J.\ would like to thank C. Klim\v c\' {\i}k for inspiring
discussions and hospitality in Marseilles, and P. \v Severa
for sharing his independent insights on quantized
twisted Poisson structures. We are grateful to Y. Soibelman for clarifying
some subtle points concerning Formality.

\end{document}